\documentstyle[aps,preprint,tighten,floats,epsf,rotate]{revtex}
\let\section=\subsection  \let\subsection=\subsubsection

\def\be{\begin{equation}}
\def\ee{\end{equation}}
\def\bea{\begin{eqnarray}}
\def\eea{\end{eqnarray}}
\def\Bphi{\mbox{\boldmath $\Phi$}}
\def\hphi{\mbox{\boldmath $\hat\Phi$}}
\def\vx{\mbox{\boldmath $x$}}

\begin{document}

%%%%%%%%%%%%%%%%%%%%%%%%%%%%%%%%%%%%%%%%%%%%%%%%%%%%%%%%%%%%%%%%%%%%%%%%%%%%%%

\preprint{\vbox{\it 
                        \null\hfill\rm    SI-TH-99-3,hep-ph/yymmnn}\\\\} 
%%%%%%%%%%%%%%%%%%%%%%%%%%%%%%%%%%%%%%%%%%%%%%%%%%%%%%%%%%%%%%%%%%%%%%%%%%%%%%

%
\title{ Skyrmions and Bags in the $2D$-$O(3)$ model }
\author{G. Holzwarth\thanks{%
e-mail: holzwarth@physik.uni-siegen.de}}
\address{Fachbereich Physik, Universit\"{a}t Siegen, 
D-57068 Siegen, Germany} 
\date{May 1999}
\maketitle

\begin{abstract}\noindent
Localized static solutions of the $2D$-$O(3)$ model are investigated
in a representation with the 3-vector field $\Bphi$ split into the unit
vector $\hphi$ and the modulus $\Phi$. As in the nonlinear version of
the model this allows for the definition
of a topological winding number $B$, and for the separation of the 
complete configuration space into distinct $B$-sectors. For small 
values of the $\Phi^4$-coupling strength the stable energy minima
in these sectors are characterized by bag formation in the modulus
field which in the standard cartesian representation
of the linear $O(3)$ model would be unstable towards decay into the
trivial $B=0$ vacuum. Stabilized by $B$-conservation they exhibit
a surprising variety of very appealing features for multiply charged
systems. With the total charge bound into one common deep bag 
opposite ways of distributing the topological charge density inside 
the bag can be realized: Pointlike structures which retain the 
individuality of single constituents (or doubly charged pairs), 
or a deconfined charge density spread uniformly throughout the
interior of the bag. 
It is suggested that this extension supplies a crucial link to 
overcome the unsatisfactory existing mismatch between 
multiskyrmion configurations and nuclear structure.
\end{abstract} 

\vspace{3cm}
\leftline{PACS numbers: 05.45.-a,11.27.+d,12.39.Dc,73.40.Hm }  
\leftline{Keywords: Sigma models, Skyrmions, Bags, Quantum Hall Effect}

\newpage

\section{Introduction}
The linear $2D$-$O(3)$ model has been a favorite tool as a model field
theory for a wide spectrum of physical systems. Commonly, the field vector
$\Bphi$ is embedded into a euclidean manifold  and
parametrized in terms of three cartesian components $\Phi_i$, (i=1,2,3).
Topologically this manifold is trivially connected. 
For standard $\Phi^4$ potentials of the type $( \sum_i \Phi_i^2 - 1)^2$
the vacuum manifold is the 2-sphere $S^2$. Addition of symmetry
breakers can remove the degeneracy of the vacuum manifold.
Imposing the constraint $\sum_i \Phi_i^2 \equiv 1$ defines the nonlinear
$2D$-$O(3)$ model.The homotopy $S^2 \to S^2$ then allows for a 
classification of static solutions by an integer topological winding
index $B$ which is the spatial integral over the time component of a
conserved topological current. Nontrivial configurations with $B \neq
0$ have found a direct application for the interpretation of the
Quantum Hall Effect as charged excitations in
two-dimensional spin systems~\cite{QHE}. In the linear model, where
configurations are not confined to the 2-sphere $S^2$, the winding
number $B$ is not topologically protected.  Therefore, beyond a
critical magnitude for the coupling constants of symmetry breakers,
static solutions with non vanishing winding $B \neq 0$  become unstable,
and they unwind into the topologically trivial ($B=0$) uniform vacuum. In
time-dependent evolutions changes in $B$ will frequently occur whenever 
the field passes through $\Bphi=0$ at some point $\vx,t$ in space  and time.

In certain applications of the model the length of the $\Bphi$-vector
is an important degree of freedom such that the constraint to $S^2$ is
too restrictive, but still, the
winding number may correspond to a physically observable conserved
charge. A prominent example is the chiral phase transition where the 
order parameter is the chiral meson field, its length measures the
amount of spontaneous symmetry breaking, while the Skyrme-Witten
conjecture~\cite{SkyWi} identifies the winding index with baryon number. Its
conservation law should not be affected by the chiral transition.
Then, in order to retain the model as a faithful
image of the underlying physics, it is necessary to prevent unwinding
of nontrivial configurations. To achieve this it is sufficient to
exclude one poimt from the manifold on which the fields live.
The natural choice is to remove the origin $\Bphi=0$, which is 
automatically accomplished in the radial-angular representation
$\Bphi=\Phi \hphi$ because the angular fields are not defined at
$\Phi=0$. Then the manifold is nontrivially connected and the winding
number is topologically protected. Thus, choosing the appropriate
embedding for the field is an important part of the definition of the
model.

In lattice simulations the topological arguments based on continuity
cannot be used because field configurations may differ arbitrarily
between two neighbouring lattice points.
However, the conservation law is easily implemented into the
update-algorithm, by allowing only for configurations with specified
value of $B$. Evidently, for that purpose it is necessary to be
able to evaluate the winding number for each configuration, so it
is necessary to work in the representation $\Bphi=\Phi \hphi$, where 
the winding density is defined in terms of angles which may vary
arbitrarily from one lattice point to its neighbours. 

So, in the following, we shall be dealing with all degrees of
freedom of the linear $O(3)$ model,
however in the form of a nonlinear $O(3)$ model supplemented by an
additional modulus field $\Phi$.
In this framework it is possible to define and conserve the
winding number of nontrivial configurations. Naturally, in a sector
with specified $B \neq 0$, we will find static solutions which do not
exist in the topologically trivial cartesian embedding
of the linear $O(3)$ model and which are characterized by the formation
of a localized spatial bag in the modulus field.

We shall discuss two versions of the model which differ in the 
definition of the current that enters the current-current coupling
Skyrme term. The competition between a symmetry-breaking Zeeman term
and the Skyrme term determines the spatial extent of the soliton
solutions. Thus the two different versions will lead to a
characteristic difference in the spatial structure of the resulting
winding density distributions in the interior of the bags:
A dominating Skyrme term deconfines the topological charge inside the
bag while a dominating Zeeman term preserves the particle nature of
individual charge units.

\section{ The $2D$-$O(3)$ model with Skyrme-Zeeman stabilization }

We consider the $O(3)$ lagrangian density in $2+1$ dimensions
in terms of the dimensionless $3$-component field
$\Bphi = \Phi \hphi$ with $\hphi \cdot \hphi =1$,
\be\label{lag}
{\cal{L}} = F^2 \left(\frac{1}{2} \partial_\mu \Bphi \partial^\mu \Bphi
- \frac{\lambda}{4\ell^2}\left( \Phi^2-f^2\right)^2
-  \frac{\alpha}{\ell^2} (f_0-\Phi_3)
-(\alpha \ell^2) \rho_\mu \rho^\mu \right).  
\ee
Apart from the usual kinetic term this lagrangian contains the
standard $\Phi^4$ potential for the modulus field $\Phi$ to monitor the
spontaneous symmetry breaking with (dimensionless) coupling parameter
$\lambda$, an explicitly symmetry-breaking ('Zeeman') coupling with
(dimensionless) strength $\alpha$, and a four-derivative ('Skyrme')
current-current coupling $\rho_\mu\rho^\mu$ for the conserved
topological current 
\be\label{top}
\rho^\mu = \frac{1}{8 \pi} \epsilon^{\mu \nu \rho} \hphi \cdot
( \partial_\nu \hphi \times \partial_\rho \hphi ) ,
\ee
satisfying $\partial_\mu \rho^\mu = 0$. 
The parameter $F^2$ sets the overall energy scale, the length $\ell$ may be
absorbed into the space-time coordinates, so it sets the size of 
localized static solutions. In order to keep the uniform minimum of the
potential for finite $\alpha$ at the vacuum value
$\Phi = f_0$  we define
\be
f^2=f_0^2-\frac{\alpha}{\lambda f_0},
\ee
then we have the $\alpha$-independent boundary condition for the modulus
$\Phi$ at spatial infinity $\Phi \to f_0$ for $r \to \infty$.
Of course, we are free to insert additional powers of the modulus field
$\Phi$ into the Skyrme and the Zeeman term, the above choice being
motivated to minimize interference with the $\Phi^4$ spontaneous 
symmetry-breaking mechanism. This choice
implies that as $f_0$ goes to zero (e.g. with increasing temperature)
the typical size $\ell$ of static defects grows like $f_0^{-1/4}$.
This may be physically not unreasonable (cf. e.g. the discussion in
the 3-dimensional case in \cite{Hans97}). (We shall discuss a different
natural choice in the next section). Having fixed the 
$\Phi$-dependence of the lagrangian as given in (\ref{lag}) and
(\ref{top}) we conveniently redefine the field and the parameters by 
\be
\tilde \Phi =\Phi f_0^{-1}, \qquad \tilde F^2=F^2 f_0^2, \qquad
\tilde \ell=\ell f_0^{-1/4}, \nonumber
\ee
\be
\tilde \lambda=\lambda f_0^{3/2}, \qquad
\tilde \alpha=\alpha f_0^{-3/2}, \qquad
\tilde f^2 = 1 - \frac{\tilde \alpha}{\tilde \lambda},
\ee
omit the tildes in the following and absorb the $\ell$'s 
into the length scale of space-time. Then we finally have
\be\label{lag1}
{\cal{L}} = F^2 \left(\frac{1}{2} \partial_\mu \Bphi \partial^\mu \Bphi
- \frac{\lambda}{4}\left( \Phi^2-1+\frac{\alpha}{\lambda} \right)^2
-\alpha \rho_\mu \rho^\mu
- \alpha (1-\Phi_3) +\frac{\alpha^2}{4 \lambda}\right).  
\ee
together with the $\alpha$-independent boundary condition 
$\Phi \to 1$ for $r \to \infty$. The constant $\alpha^2/(4 \lambda)$  
has been added to set the $\Phi$-potential to zero at the 
minimum $\Bphi=1$.
The {\it nonlinear} $O(3)$-model emerges in the limit $\lambda
\to\infty$ where $\Bphi$ is confined to the 2-sphere
$\Bphi^2 \equiv 1$. For this case the static solutions for the angular
field $\hphi$ 
have been thoroughly investigated~\cite{Durham}. They are
characterized by fixed integer winding number $B$, and by a 
modulus field $\Phi(\vx)$
which for very large $\lambda$ differs only minimally from its vacuum
value $\Phi=1$. 
We shall in the following denote them as $B$-skyrmions. 

For fixed $\alpha \ne 0$ and sufficiently small values of $\lambda$
the point $\Bphi = 1$ is the only real minimum of the potential in 
(\ref{lag1}) , i.e. for $\lambda$ smaller than some critical value
$\lambda_c$ static solutions in the cartesian representation 
of the {\it linear} 
$O(3)$-model collapse to the trivial vacuum $\Bphi \equiv 1$ . In a 
$\lambda$-$\alpha$ diagram the line $\lambda_c(\alpha)$ separates two
phase regions where for $\lambda > \lambda_c(\alpha)$ static
multi-skyrmions with winding number $B$  can exist at local energy minima $E_B$,
while for $\lambda < \lambda_c(\alpha)$ only the global minimum 
$\Bphi \equiv 1$ at $E_0=0$ survives. There will be different
phase boundaries $\lambda_c(\alpha)$ for multi-skyrmions with different
winding numbers $B$. 

In the $\Bphi = \Phi \hphi$ representation of the linear $O(3)$ model
the winding number is fixed (and can be held fixed in lattice 
simulations). This means that for $\lambda > \lambda_c(\alpha)$ 
the local (multi-skyrmion) minimum $E_B$ 
turns into a global minimum in the respective $B$-sector
and for $\lambda < \lambda_c(\alpha)$ the collapse of the
multi-skyrmions into the vacuum is prevented. Instead, the global minimum
in a given $B$-sector continues to exist for decreasing values of
$\lambda$. But we expect a qualitative change in the structure 
of stable static configurations of (\ref{lag1}) in the vicinity of the
phase boundaries $\lambda_c(\alpha)$ leading to new types of solutions for 
$\lambda < \lambda_c$ which do not exist in the cartesian embedding of
the linear $O(3)$-model. The nature of
this structural change is easily visualized: Destabilization
and unwinding can only proceed through the field configuration passing 
through $\Phi=0$ at some point. Conservation of winding number
therefore leads to formation of a spatial bag: In its interior the modulus
field $\Phi$ is close to zero, with rapid variation of the angular
fields $\hphi$ such that the winding density is located inside the bag.
Thus we may expect a close relation between the deviation of the scalar
modulus field from its background value $\Phi=1$ and the density of the
topological charge. We shall in the following denote these types of
solutions as $B$-bags.

\section{Bag formation}

\begin{figure}[h]
\begin{center}
\leavevmode
\vbox{\epsfxsize=10truecm \epsfbox{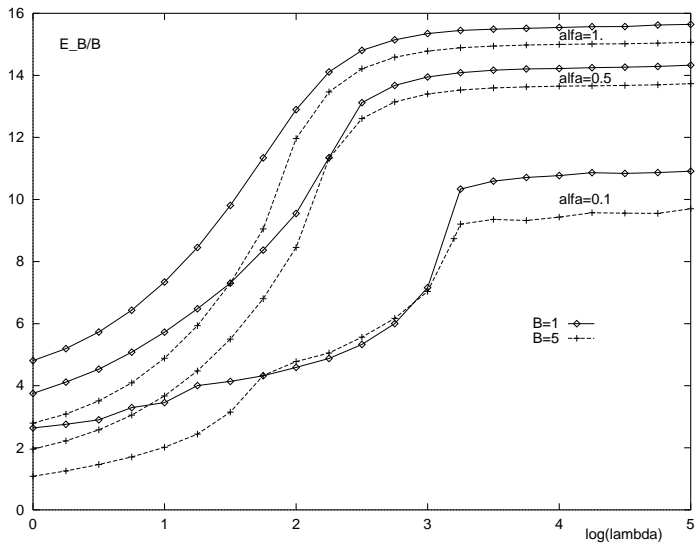}}
\end{center}
\caption{
Minimal energies per unit topological charge 
$E_B(\alpha,\lambda)/B$ as
function of $\log\lambda$ for three different values of $\alpha=0.1,0.5,1.0$.
Full and dashed lines show the cases $B=1$ and $B=5$, respectively.}
\label{fig1} 
\end{figure}

Within a given $B$-sector the transition from skyrmions to bags is
smooth. As $\lambda$ approaches $\lambda_c$ from
above creation of the bag begins through 
formation of a shallow depression in the modulus $\Phi$ near the
center of the skyrmions. As $\lambda$ passes the critical value the
depression quickly deepens with the winding density accumulating inside.
In fig.~\ref{fig1} this transition is illustrated in the $B=1$ and $B=5$
sectors. Plotted is the minimal energy per unit topological charge 
$E_B(\alpha,\lambda)/B$ as
function of $\lambda$ for three different values of $\alpha$.
The full lines show the $B=1$ sector. For a small value of $\alpha=0.1$
the transition near $\log \lambda \approx 3.2$ is most pronounced.
Bag formation sets in rather sharply, with a sudden steep decrease in
$E_B$ with decreasing $\lambda$. For very large $\lambda$ above this
critical value, $E_B$ rises very slowly towards a limit slightly above
the Belavin-Polyakov (BP) monopole energy of $4\pi$~\cite{BP}. For
larger values of $\alpha$ the transition region gets smoothed out,
still with a large gain in energy through formation of the bag.

%%%%%%%%%%%%%%%%%%%%%%%%%%%%%%%%%%%%%%%%%%%%%%%%%%%%%%%%%%%%%%%%
%%%%%%%%%%%%%%%%%%%%%%%%%%%%%%%%%%%%%%%%%%%%%%%%%%%%%%%%%%%%%%%%%%

The dashed lines show the same features in the $B=5$ sector.
Comparing the energies for $B=1$ and $B=5$ for small $\lambda$ values near
$\lambda \approx 1$ which correspond to well-developped deep bags we
see that $E_5 < 5 E_1$ for all three values of $\alpha$.
This implies that the five topological charges are strongly bound in the
compact $B=5$ bag. It is only for the smallest value of $\alpha$
considered that with increasing $\lambda$ near $\log \lambda \approx
1.75$ the 5-bag breaks up into five individual 1-bags such that $E_5/5$
closely follows $E_1$ into the transition region where the bags
disappear and the emerging five 1-skyrmions combine into pairs of two
2-skyrmions plus one left-over 1-skyrmion (for $\log \lambda > 3.2$).
For the larger $\alpha$-values considered, $E_5/5$ always stays below
$E_1$, so at no point there is a complete breakup into five individual
1-bags. 

%%%%%%%%%%%%%%%%%%%%%%%%%%%%%%%%%%%%%%%%%%%%%%%%%%%%%%%%%%%%%%%%
\begin{figure}[h]
\begin{center}
\leavevmode
\vbox{\epsfxsize=10truecm \epsfbox{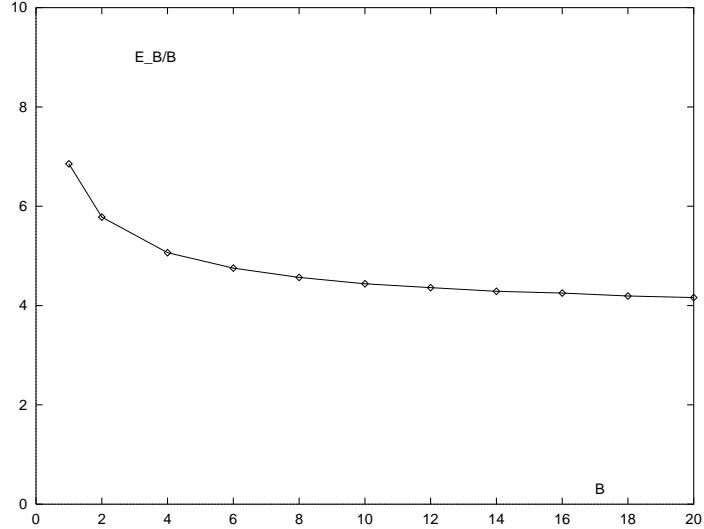}}
\end{center}
\caption{ Minimal energy $E_B$ per unit topological charge $B$ 
 as function of $B$. ($\lambda=10$, $\alpha=1$). }
\label{fig2} 
\end{figure}
%%%%%%%%%%%%%%%%%%%%%%%%%%%%%%%%%%%%%%%%%%%%%%%%%%%%%%%%%%%%%%%%%%

This holds for all values of $B$ for sufficiently small $\lambda$.
Figure \ref{fig2} shows the energy per unit topological charge at
$\lambda=10$ and 
$\alpha=1$. It is a monotonically decreasing function of $B$
which converges towards approximately one half of 
the $B=1$ energy; so for this $\lambda$ all B-bags are stable against
decay into 1-bags. 
Altogether there emerges a phase diagram of remarkable richness.
The skyrmion regions ($\lambda \gg 1$) have been amply discussed, 
so here we only illustrate typical features of the region with well
developped deep bags. (In order to get sufficient resolution the results
are calculated for $\ell=10$ on an 80$\times$80 mesh.)

%%%%%%%%%%%%%%%%%%%%%%%%%%%%%%%%%%%%%%%%%%%%%%%%%%%%%%%%%%%%%%%%
\begin{figure}[h]
\begin{center}
\leavevmode
\hbox{\epsfysize=6truecm \epsfbox{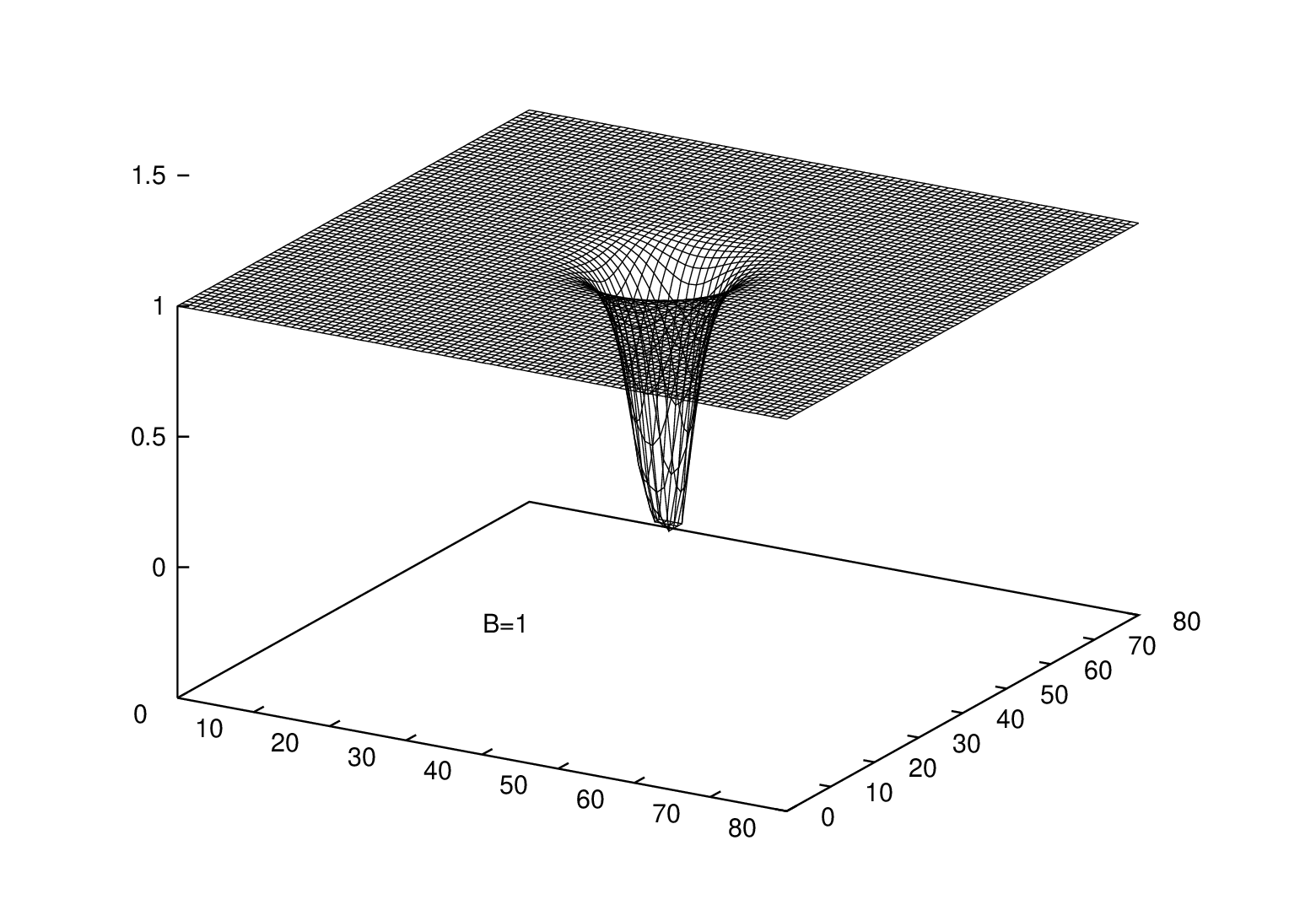} 
\epsfysize=6truecm \epsfbox{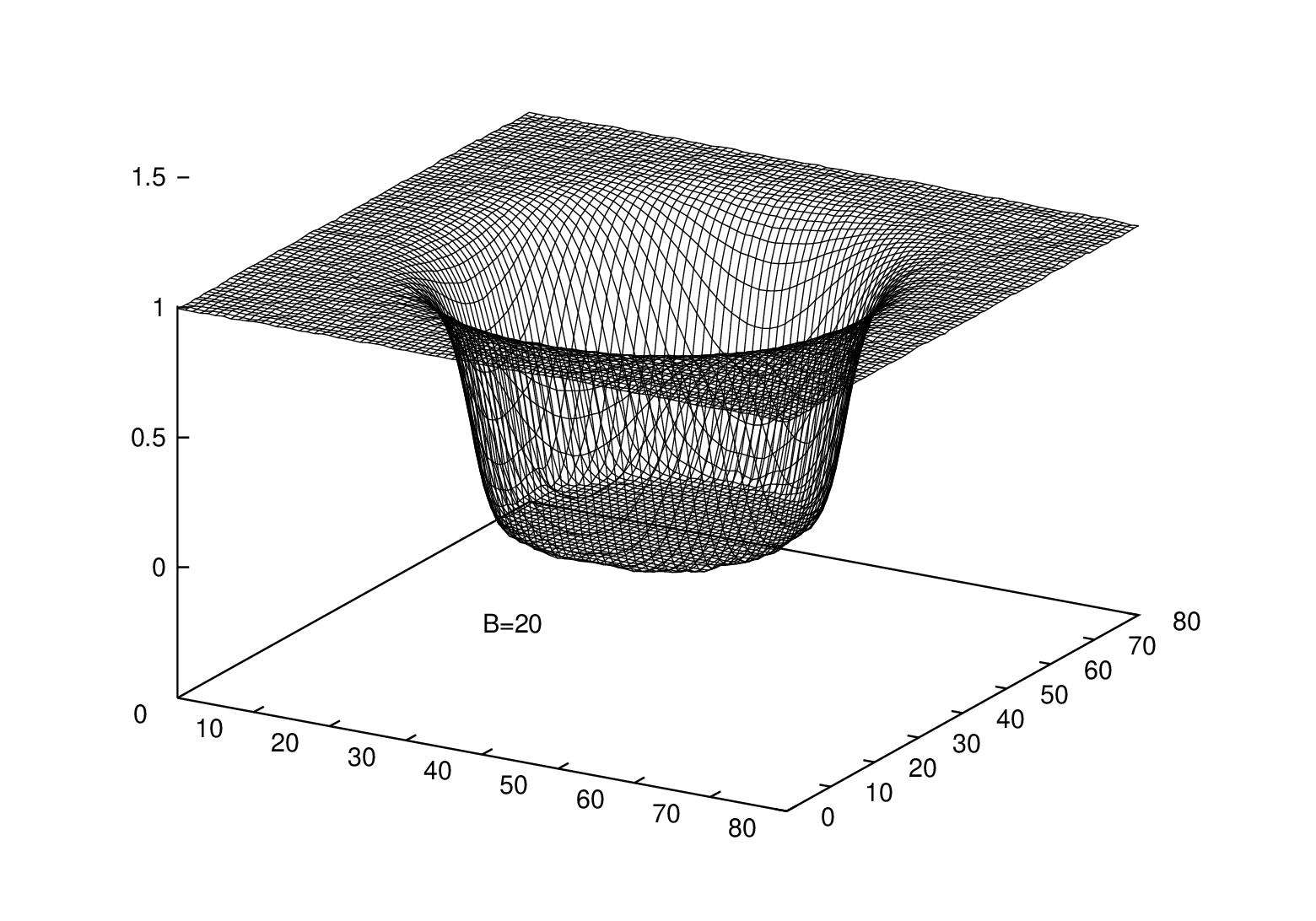}}
\end{center}
\caption{
3D-view of the bags formed for $B=1$ (a), and $B=20$ (b), (for $\alpha=1$,
$\lambda=10$). }
\label{fig3} 
\end{figure}
%%%%%%%%%%%%%%%%%%%%%%%%%%%%%%%%%%%%%%%%%%%%%%%%%%%%%%%%%%%%%%%%%%
%%%%%%%%%%%%%%%%%%%%%%%%%%%%%%%%%%%%%%%%%%%%%%%%%%%%%%%%%%%%%%%%
\begin{figure}[h]
\begin{center}
\leavevmode
\vbox{\epsfxsize=11truecm 
\epsfysize=6truecm \epsfbox{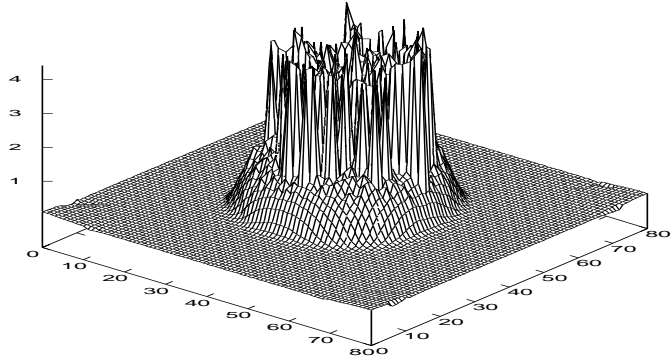}}
\end{center}
\caption{ 3D-view of the angular field $\Theta(\vx) = \arccos
\hat\Phi_3(\vx)$ for $B=20$ ($\lambda=10$, $\alpha=1$),
fluctuating rapidly inside the bag of fig.~\ref{fig3}b around an
average value of $\pi$. }
\label{fig4} 
\end{figure}
%%%%%%%%%%%%%%%%%%%%%%%%%%%%%%%%%%%%%%%%%%%%%%%%%%%%%%%%%%%%%%%%%%

%%%%%%%%%%%%%%%%%%%%%%%%%%%%%%%%%%%%%%%%%%%%%%%%%%%%%%%%%%%%%%%%
\begin{figure}[h]
\begin{center}
\leavevmode
\hbox{\epsfysize=6truecm \epsfbox{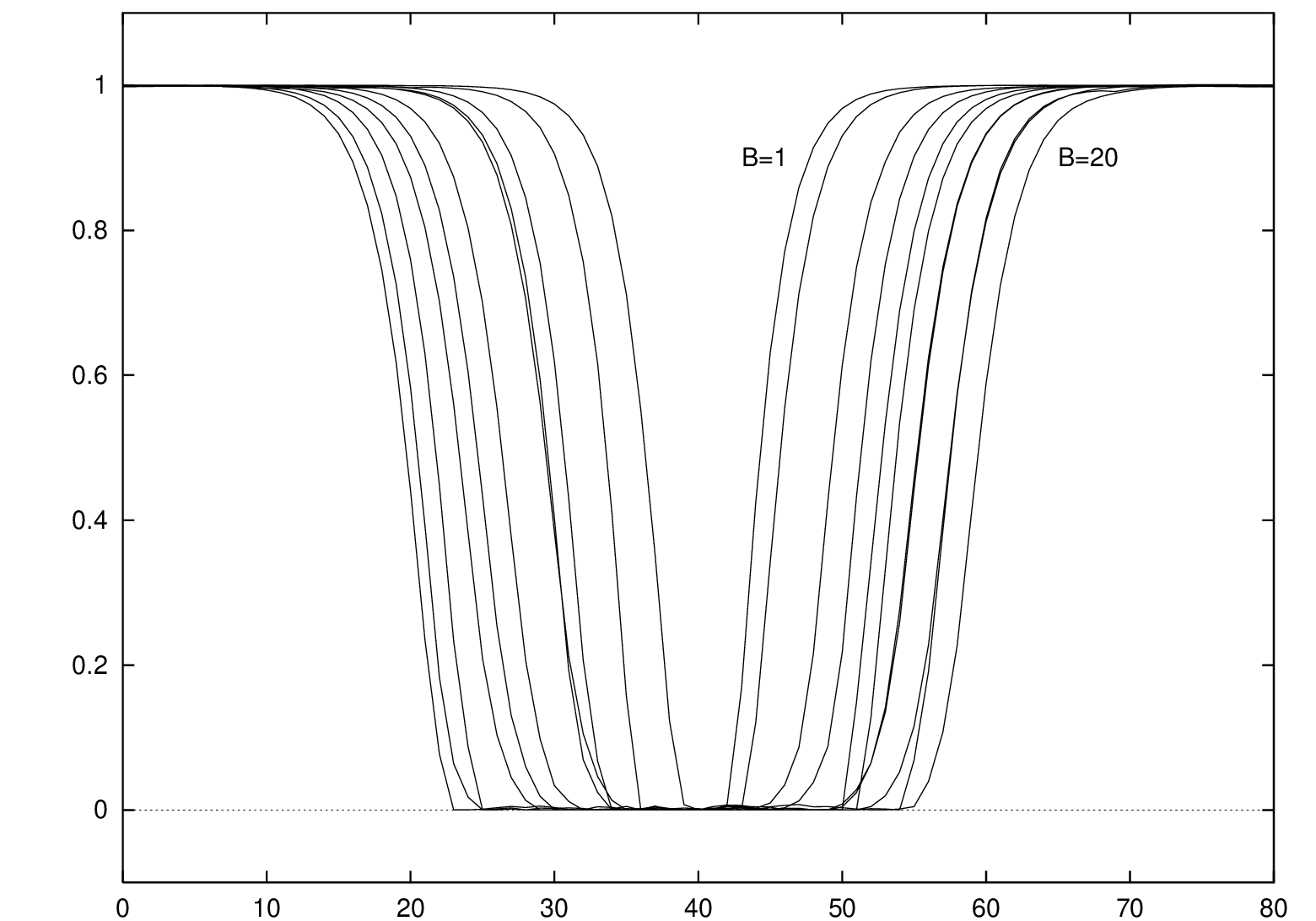} 
\epsfysize=6truecm \epsfbox{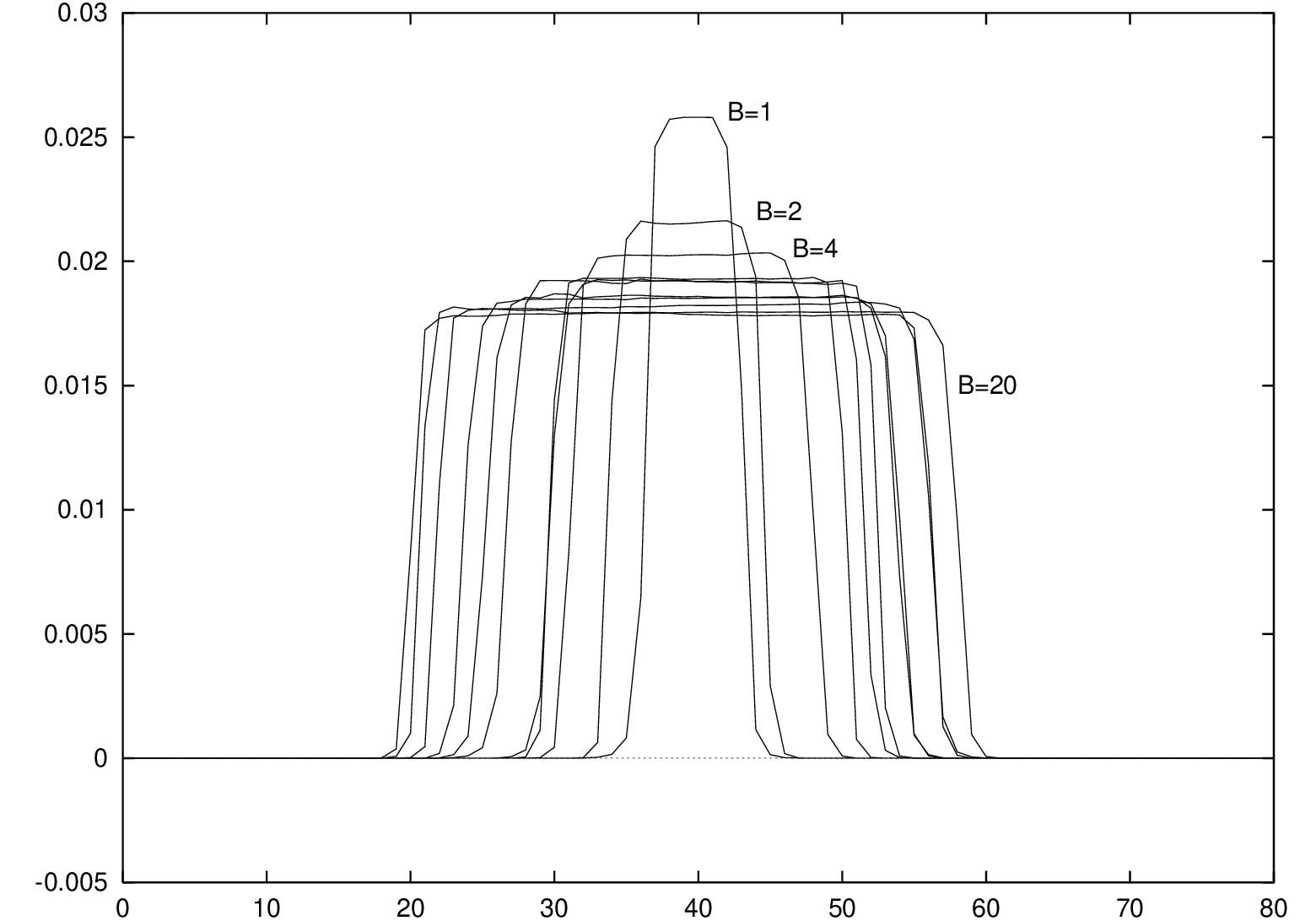}}
\end{center}
\caption{ Near-central cuts through the bag profiles (a), and the
corresponding winding density profiles (b), 
(for $B$=1,2,4,6,..,20) for $\lambda=10$, $\alpha=1$. 
(Due to numerical fluctuations the positions of the bag
centers may differ slightly for different values of $B$.)
}
\label{fig5} 
\end{figure}
%%%%%%%%%%%%%%%%%%%%%%%%%%%%%%%%%%%%%%%%%%%%%%%%%%%%%%%%%%%%%%%%%%
The 1-bag which is formed for $\alpha=1$ and $\lambda=10$ is
shown in fig.~\ref{fig3}a. Its total classical energy is $E_1=6.85$.  
Already for this smallest value of $B$ the bag has developped a
basically flat bottom where the modulus field is very close to zero,
with small numerical fluctuations. 
To accommodate increasing numbers of topological charge $B$ the bag
size increases correspondingly, its depth being limited by
$\Phi>0$. As an example, fig.\ref{fig3}b shows the case $B=20$ for
$\lambda=10$ and $\alpha=1$. 
The near-central cuts through the bag profiles plotted in
fig.~\ref{fig5}a (for $B$=1,2,4,6,..,20) show that the surface
thickness of the 
bags is basically independent of $B$, it is only the radius of the
flat interior which adjusts to the increasing total charge.

Throughout this deep bottom of the bag the angular fields display
numerous strong local fluctuations. 
(Fig.\ref{fig4} shows the angle $\Theta = \arccos \hat\Phi_3$ (for $B=20$)
fluctuating rapidly inside the bag around an average value of
$\pi$). Such fluctuations are energetically harmless because the large
gradient terms are multiplied by the almost vanishing square of the 
modulus field. By this mechanism an essentially constant winding
density $\rho_0$ is achieved which extends over the whole flat interior
of the bag (apart from small numerical fluctuations, see fig.\ref{fig5}b). 
In the context of nuclear physics one would say
that the nucleons have lost their identity inside the nucleus and have
dissolved into a hadronic soup. 
In fig.\ref{fig5}b the near-central cuts through the density profiles
are shown which correspond to the same bags as given in
fig.\ref{fig5}a. Again the surface 
thickness of the profiles is basically constant; it is, however,
smaller than that of the bag profiles. So, for larger values of $B$
a square slab closely approximates the density distributions. 
This is to be contrasted with the winding density of a B-skyrmion. 
Its angular fields closely resemble the BP monopole
$\Theta(r) = 2 \arctan(B/r)$ with winding density
\be
\rho_{BP}=-\frac{B \Theta' \sin \Theta}{4 \pi r}=\frac{B^3}{\pi
(r^2+B^2)^2}.
\ee
There is no central plateau, the center density decreases as
$\rho_{BP}(0)\propto B^{-1}$ and the mean radius increases as 
$R\propto B$.
In contrast to this, fig.\ref{fig5}b shows that for the B-bags the central
density $\rho_0$ quickly drops from its 
maximal value for $B=1$ ($\rho_0=0.025$ for $\lambda=10, \alpha=1$),
then slowly converges with
increasing $B$ towards a constant value, which for  
this parameter set is near $\rho_0 \approx 0.017$. 
This $B$-dependence reflects the interplay between the
surface energy of the bag (which originates in the gradient terms of
the modulus field) and volume energies which comprise the potential
energy of the modulus field, Zeeman and Skyrme terms, and kinetic terms
due to the gradients of the angular fields in the linear $\sigma$-model
part of the lagrangian (\ref{lag1}).

To get an idea about the ('nuclear matter') density $\rho_0$
in the limit $B\to\infty$ (where surface terms play no role) we could
approximately replace the winding density defined in terms of the angular
fields (\ref{top}) by the deviation 
of the modulus field $\Phi$ from its vacuum value
\be
\rho \approx \rho_0(1-\Phi)
\ee
and obtain the bag radius $R$ from $\pi R^2 \rho_0 =B$. 
The lagrangian (\ref{lag}) then shows that creation of a deep bag
($\Phi\approx 0$ inside, $\Phi$=1 outside) requires a volume potential 
energy density of 
\be
\epsilon_V = \frac{\lambda+2\alpha}{4 \ell^2}+\ell^2\alpha \rho_0^2.
\ee 
Ignoring kinetic contributions for the moment,
for fixed $B$ the interplay between these two volume terms determines
the average density $\rho_0$ as
\be
\label{dens}
\rho_0=\frac{1}{2\ell^2}\sqrt{\frac{\lambda+2\alpha}{\alpha} }.
\ee
which for $\lambda=10$ and $\alpha=1$ is $\rho_0 = \sqrt 3 /\ell^2=0.0173$.

The additional pressure due to the kinetic volume terms will further lower
the central density, but comparison with the numerical result in
fig.\ref{fig5}b shows that their influence must be small. 
This is what we might expect 
due to the smallness of $\Phi$ inside the bag, although very large
gradients in $\hphi$ could compensate for it and cause a noticeable
lowering of $\rho_0$, especially for small values of $\alpha$.
Maximal importance of kinetic terms would occur if
the hadronic soup inside the bag would consist of fermions.
Then, in Thomas-Fermi approximation for the density 
$\rho_0=p_F^2/(4 \pi)$ we would have for the energy per particle 
\be
\frac{\epsilon}{\rho_0}=\frac{4 \sqrt{\pi}}{3}\sqrt{\ell^2\rho_0}+
\frac{\epsilon_V}{\rho_0}.
\ee
Variation with respect to $\rho_0$ then leads to
\be
\label{TF}
\frac{2\sqrt{\pi}}{3}\ell\rho_0^{3/2}-
\frac{\lambda+2\alpha}{4\ell^2 } + \ell^2 \alpha \rho_0^2 =0.
\ee
For $\lambda=10$ and $\alpha=1$ the central density thereby is lowered 
from 0.0173 to $\rho_0=0.0120$. Comparison with fig.\ref{fig5}b
apparently rules out the 
conjecture of a fermionic soup. (For small values of $\alpha$ the 
effect of the Thomas-Fermi term becomes stronger: for 
$\lambda=10$, $\alpha=0.1$, $\ell=10$, it lowers
the result of (\ref{dens}) $\rho_0=0.050$ to $\rho_0=0.016$, while
the numerically obtained density for $B=20$ is near 0.04).

\section{Individual particles inside the bag}

An interesting alternative way to define the Skyrme term in the
lagrangian is to replace in (\ref{lag}) the topological current 
$\rho^\mu$ (\ref{top})
by the corresponding form in terms of the full vector field $\Bphi$
\be\label{top1}
T^\mu = \frac{1}{8 \pi} \epsilon^{\mu \nu \rho} \Bphi \cdot
( \partial_\nu \Bphi \times \partial_\rho \Bphi ) = \Phi^3 \rho^\mu.
\ee
This introduces a factor $\Phi^6$ into the Skyrme term. Due to this high
power of $\Phi$ there again is very little interference with the 
$\Phi^4$ spontaneous symmetry-breaking mechanism of the $\Phi$-potential.
However, apart from the fact that
the length $\ell$ now scales with $f_0^{5/4}$ as $f_0$ tends to zero,
this alternative definition leads to a situation where, for fixed $f_0$, in
spatial regions with small $\Phi$ (i.e. in the interior of bags)
the Zeeman term dominates the Skyrme term. This results in the stabilization
of pointlike 'particles' inside the bag. Speaking again in the language
of nuclear physics, one would say that the nucleons retain their
individuality inside a bag which binds them together into a common
$B$-nucleus.

%%%%%%%%%%%%%%%%%%%%%%%%%%%%%%%%%%%%%%%%%%%%%%%%%%%%%%%%%%%%%%%%
\begin{figure}[h]
\begin{center}
\leavevmode
\hbox{\epsfxsize=5.7truecm \epsfysize=6truecm \epsfbox{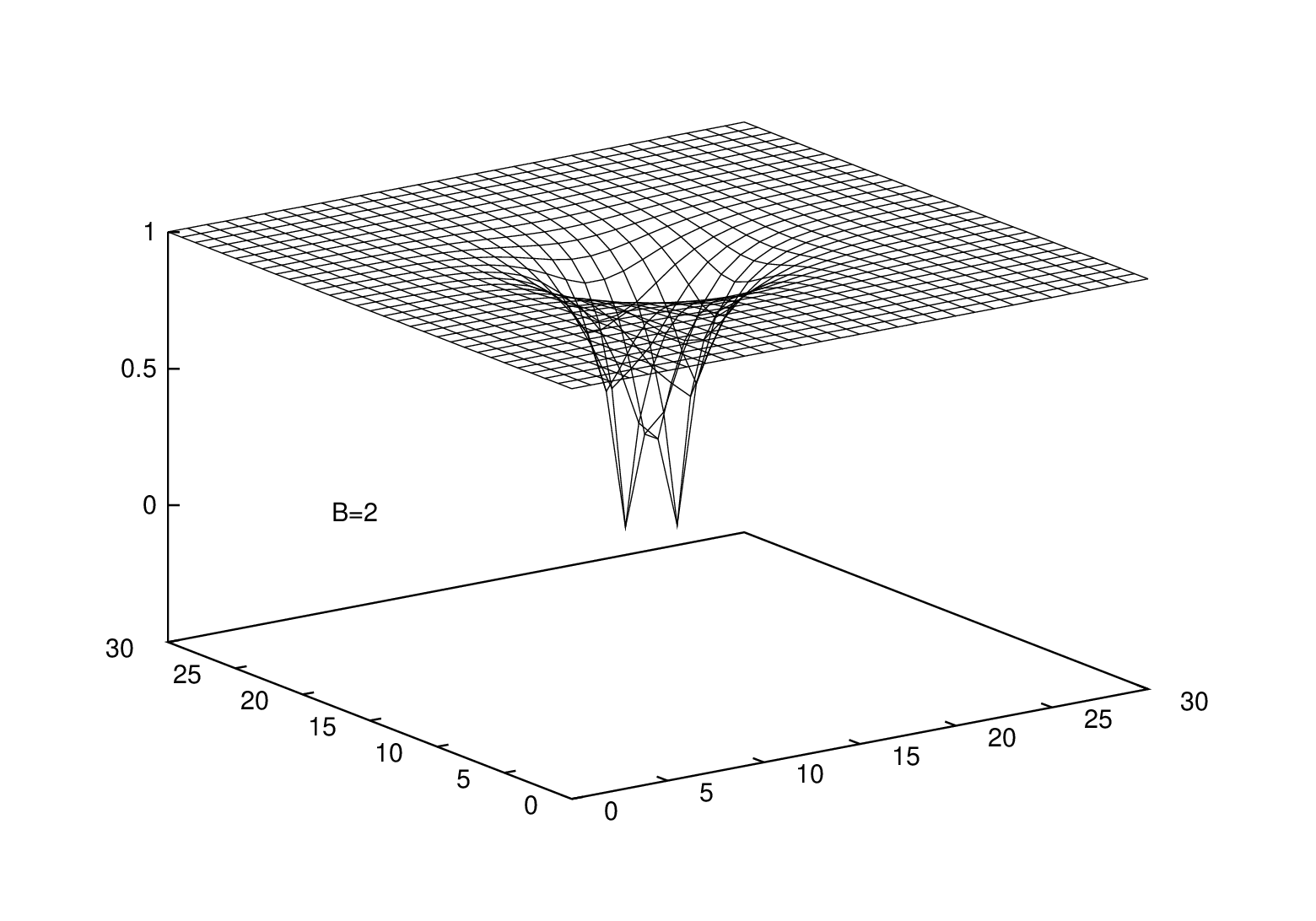} 
\epsfxsize=5.7truecm \epsfysize=6truecm \epsfbox{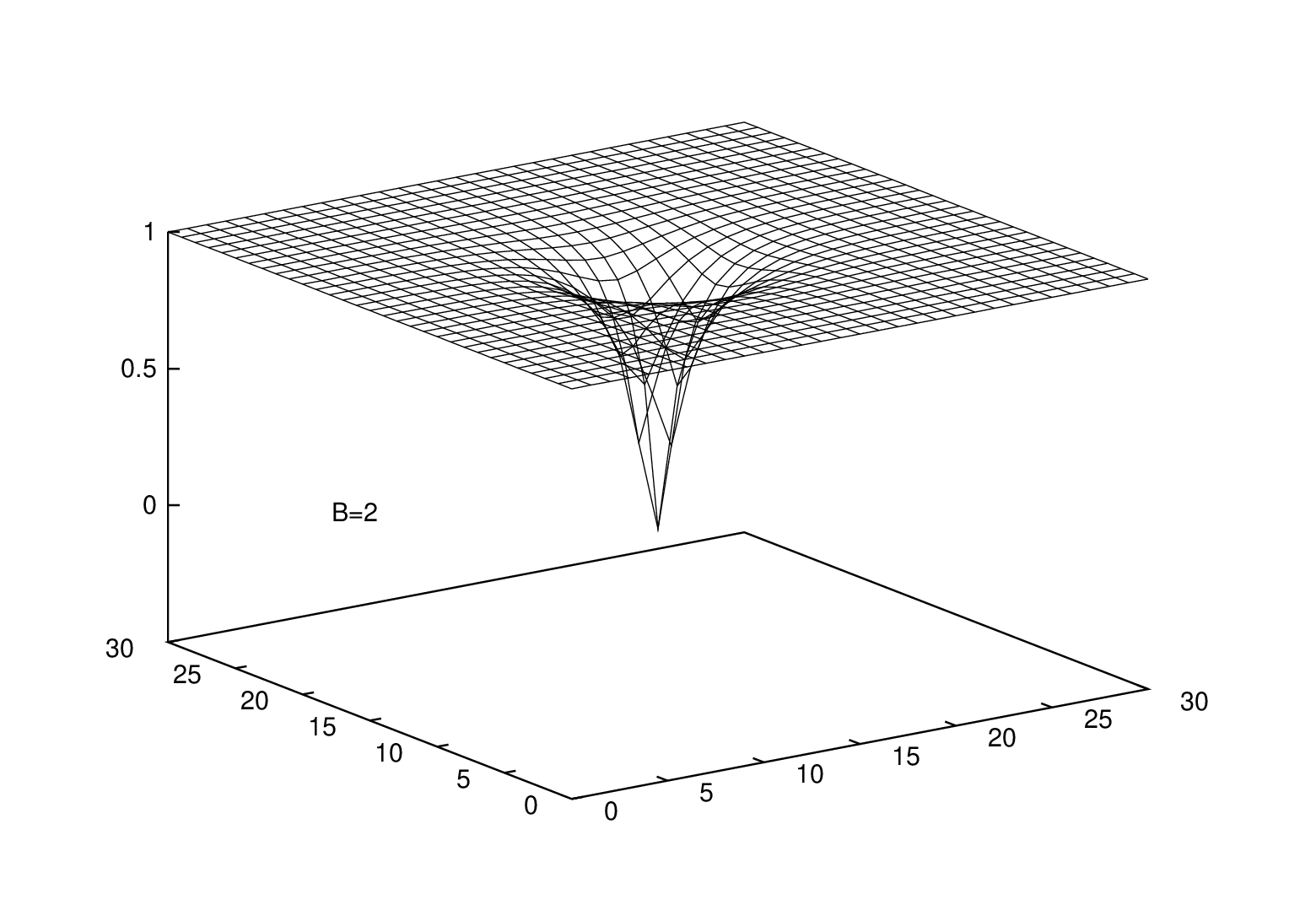}
\epsfxsize=5.7truecm \epsfysize=6truecm \epsfbox{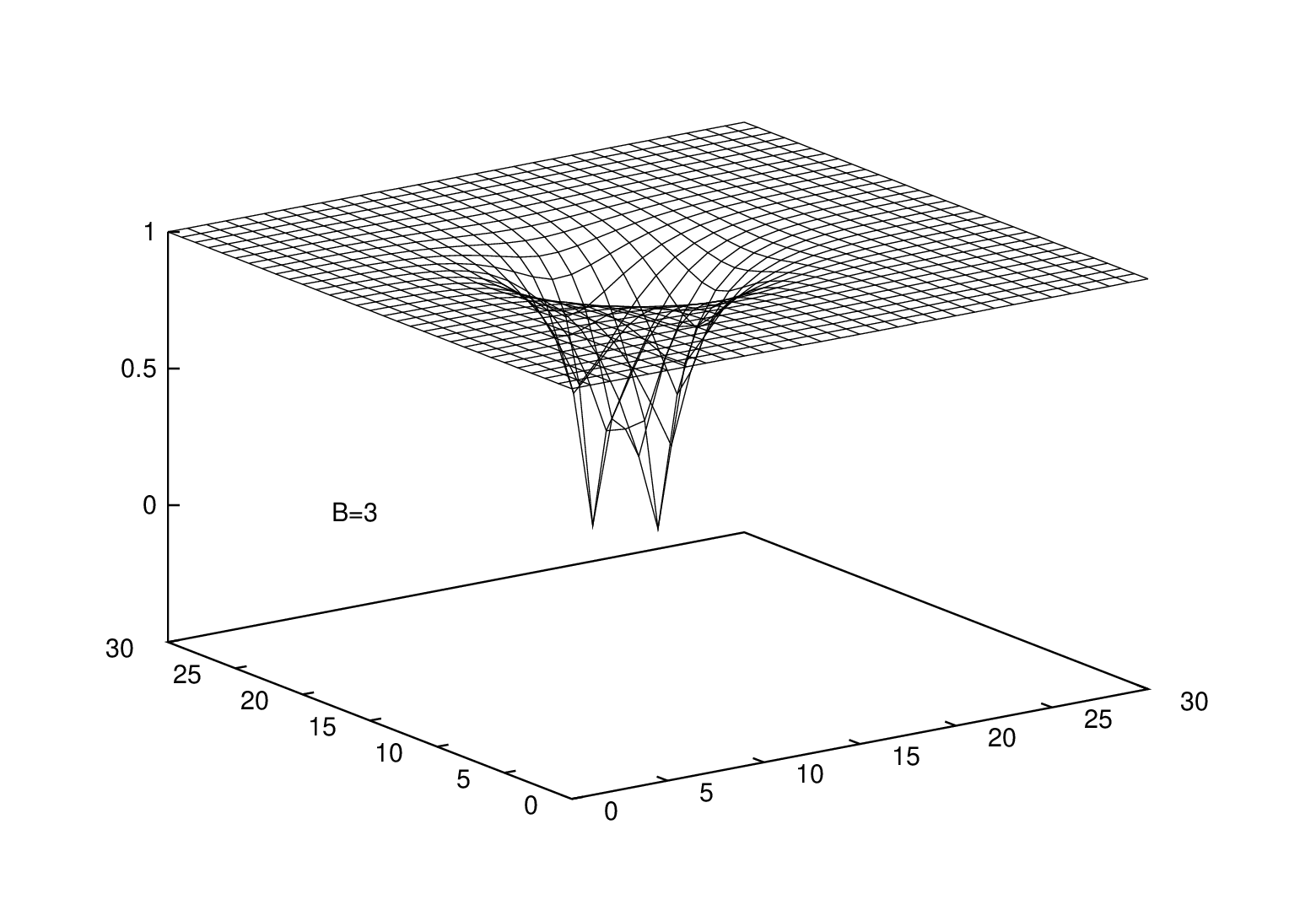}}

\hbox{\epsfxsize=5.7truecm \epsfysize=6truecm \epsfbox{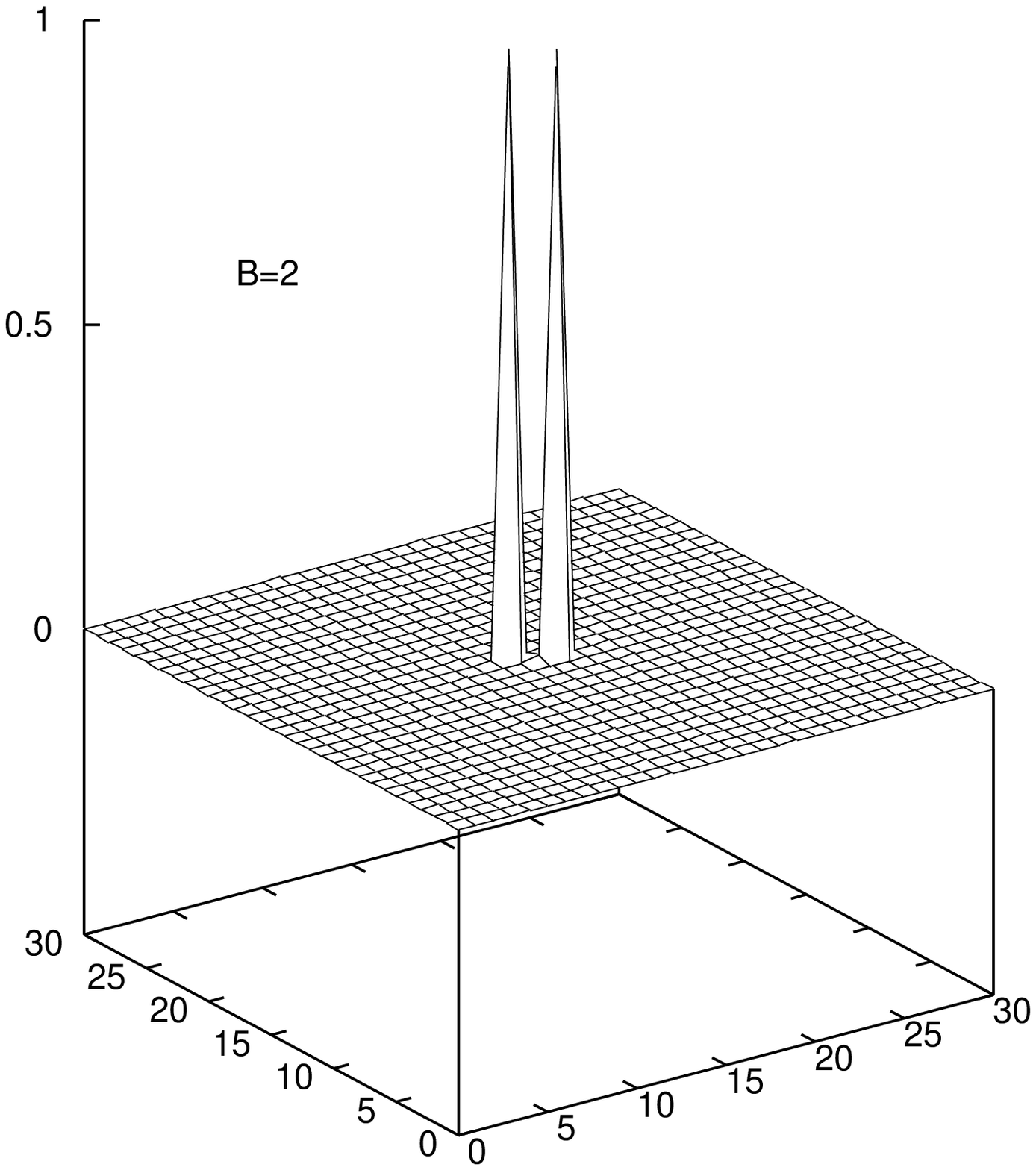} 
\epsfxsize=5.7truecm \epsfysize=6truecm \epsfbox{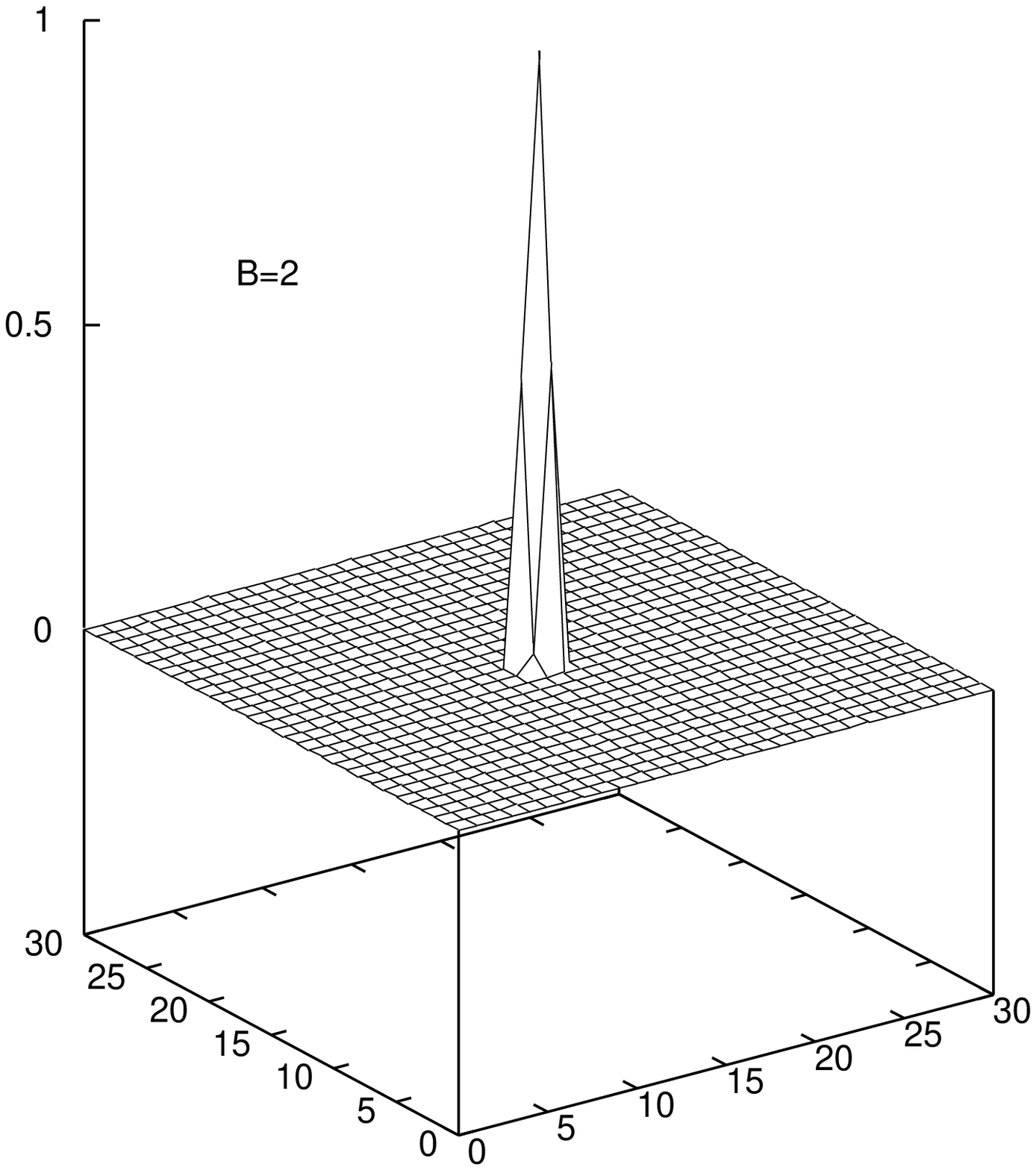}
\epsfxsize=5.7truecm \epsfysize=6truecm \epsfbox{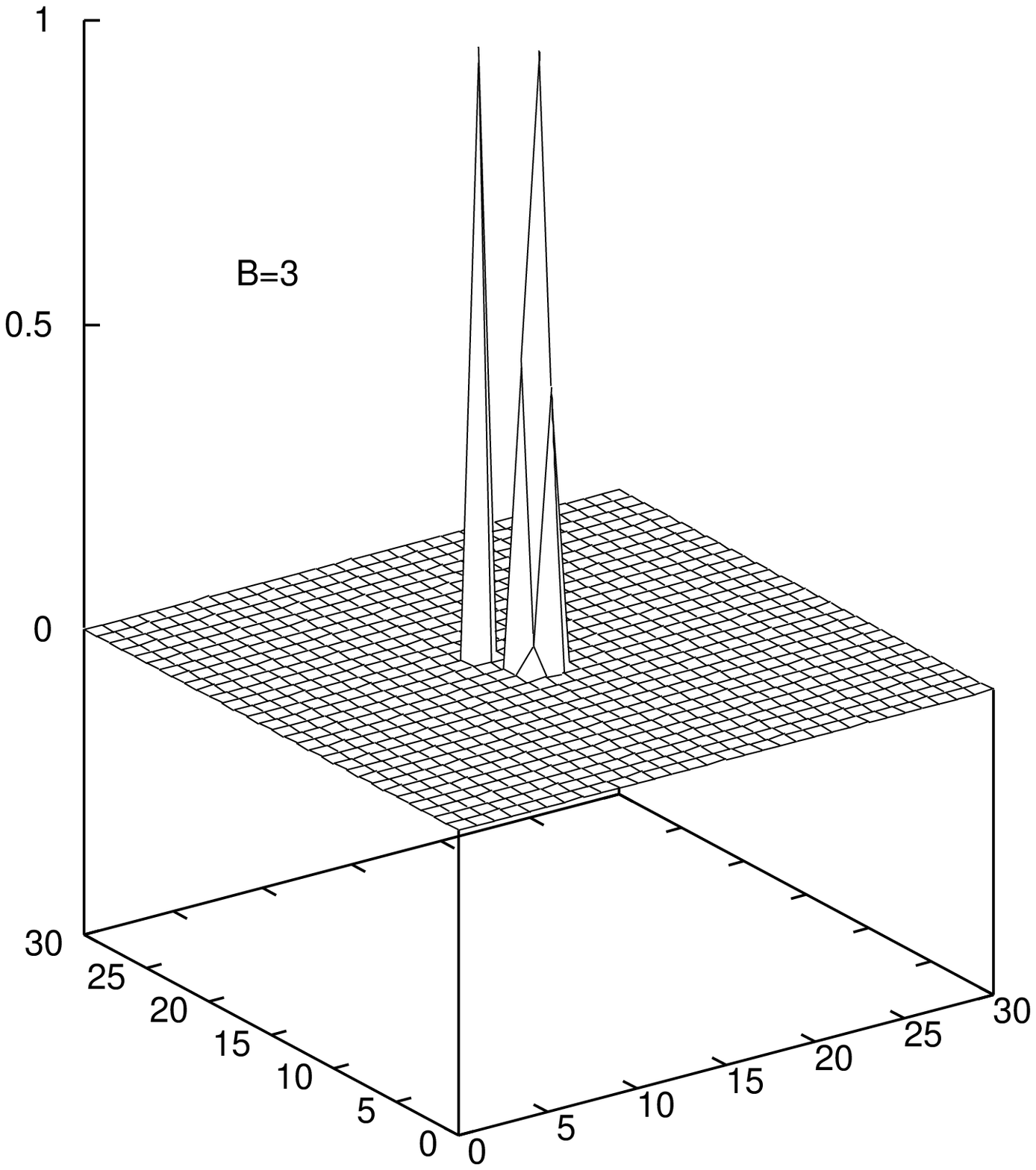}}
\end{center}
\caption{
3D-view of the bags and corresponding winding densities for the
'1+1' and '2' configurations observed in the $B=2$ sector (left and
center) and the '1+2' configuration in the $B=3$ sector (right),
if the Skyrme term in (\ref{lag}) is multiplied by $\Phi^6$ 
as follows from the alternative definition (\ref{top1}).
($\alpha=1$, $\lambda=10$). The energies per unit charge are
$E_B/B=0.943,0.663,0.697$, respectively. }
\label{fig6} 
\end{figure}
%%%%%%%%%%%%%%%%%%%%%%%%%%%%%%%%%%%%%%%%%%%%%%%%%%%%%%%%%%%%%%%%%%

%%%%%%%%%%%%%%%%%%%%%%%%%%%%%%%%%%%%%%%%%%%%%%%%%%%%%%%%%%%%%%%%
\begin{figure}[t]
\begin{center}
\leavevmode
\hbox{\epsfxsize=8.5truecm 
\epsfysize=8truecm \epsfbox{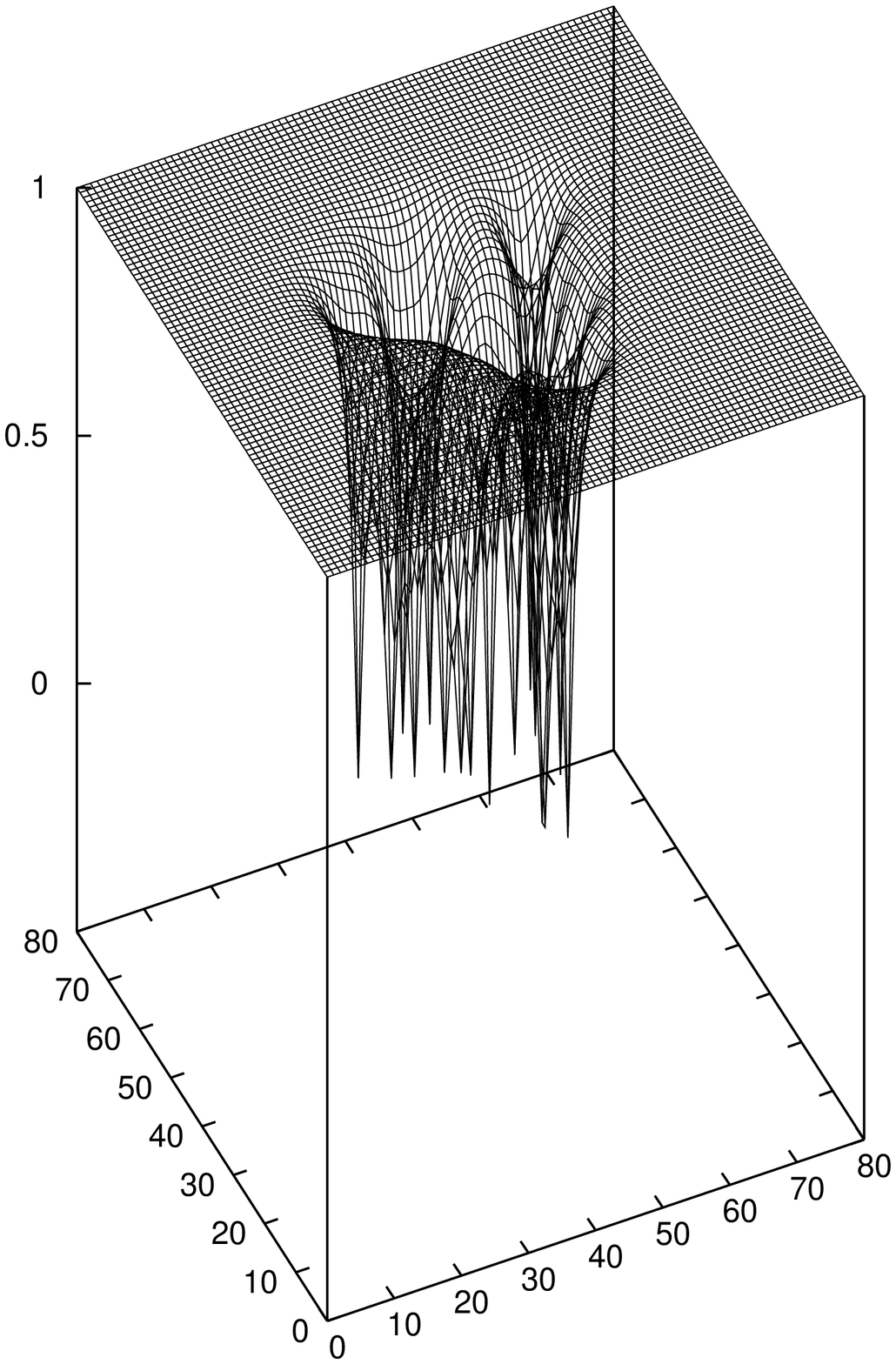} 
\epsfxsize=8.5truecm 
\epsfysize=8.8truecm \epsfbox{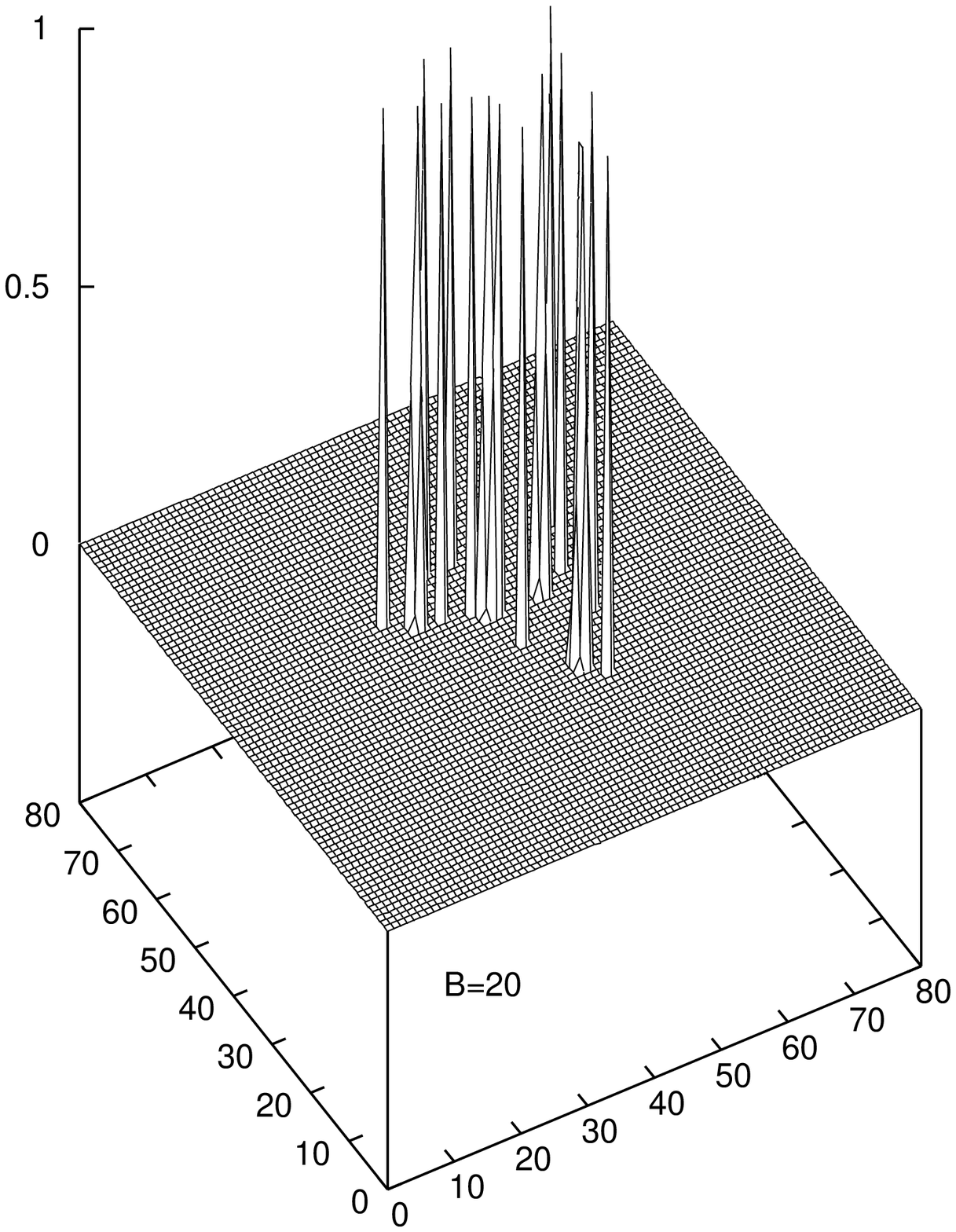}}
\end{center}
\caption{
3D-view of the bag formed for $B=20$, and the corresponding winding
density $\rho$, if the Skyrme
term in (\ref{lag}) is multiplied by $\Phi^6$ as it follows from the
alternative definition (\ref{top1}). The parameters ($\alpha=1$ and
$\lambda=10$) are the same as in fig.~\ref{fig3}b. }
\label{fig7} 
\end{figure}
%%%%%%%%%%%%%%%%%%%%%%%%%%%%%%%%%%%%%%%%%%%%%%%%%%%%%%%%%%%%%%%%%%
Choosing as before the parameters $\lambda=10,\alpha=1$ which favor the
formation of deep bags, we find the energy minimum for $B=1$ at
$E_1=1.152$. In the $B=2$ sector we find a minimum near $E_2/2=0.943$
in which both charges share one common bag, but the
individual charges stay apart from each other in two separate
pockets. This '1+1'-configuraton is shown on the left side of
fig.\ref{fig6}. There is, however, in this sector a lower minimum
at $E_2/2=0.663$ where the bag consists of only one single pocket which
houses one doubly charged structure, i.e. the model favors the
formation of pairs.  This '2'-configuration
is shown in the center plot of fig.\ref{fig6}. For $B=3$ we find the 
expected '1+2'-configuration at $E_3/3=0.697$, a pair and one unit
charge tightly bound in two separate pockets of a
common 3-bag,  as shown on the right of
fig.\ref{fig6}. We do not find a minimal-energy '3'-configuration with
a threefold charge sitting in one single pocket. The same situation is
observed in the $B=4$ sector where again we do not find the single
pocket with fourfold 
charge, but instead the '2+2' double pair separated in two pockets
bound in one common 4-bag at $E_4/4=0.543$. 
So, it appears that the situation resembles very closely the pair formation
as it was observed~\cite{Durham} for skyrmions for $\lambda \to
\infty$. As in that case, the '1'- and '2'-configurations serve as
building blocks which here are bound together in one deep bag. 
Due to this energetically favored pair
formation the resulting 'nuclei' will be characterized by
odd-even staggering of their binding energies for higher values of $B$.
Practically, of course, the actual configuration which is finally reached
through numerical relaxation depends on the starting set and the
sequence of pseudo temperatures in the Metropolis cooling algorithm. 
As an example we present
in fig.\ref{fig7} a result for $B=20$ which shows a highly deformed
bag with 15 pockets which houses a $10 \times '1'+5 \times'2'$
configuration. The calculation is done for the same parameter set as
used in fig.\ref{fig3}b, so a direct comparison shows nicely the 
contrast between pointlike particles and deconfined charge in the 
interior of one big bag.

\section{Conclusion}
We have investigated here the $2D$-$O(3)$ model in a representation
where the 3-vector field $\Bphi$ is split into the unit vector $\hphi$
and the modulus $\Phi$. This allows for the definition of a topological
winding number $B$, and for the separation of the complete
configuration space into distinct $B$-sectors. For small values of the
$\Phi^4$-coupling strength $\lambda$ the stable energy minima
in these sectors are characterized by bag formation in the modulus
field. In the standard cartesian representation of the linear $O(3)$
model such configurations would be unstable towards decay into the
trivial $B=0$ vacuum. Stabilized by $B$-conservation they exhibit
a surprising variety of very appealing physics for multiply charged
systems. For decreasing $\lambda$ multi-skyrmions 
get bound into one common deep bag like nucleons get bound
into one nucleus. Depending on the competition between Skyrme and
Zeeman energy two opposite ways of distributing the topological charge
inside the bag can be realized: Pointlike structures which keep the
individuality of single nucleons (or doubly charged pairs) inside the
nucleus, or a deconfined charge density spread uniformly throughout the
interior of the bag. This latter case suggests a very close relation
between the charge density $\rho$ and the modulus field $\Phi$ 
(which for sufficiently large $B$ is $\rho=\rho_0(1-\Phi)$)
and, correspondingly, an effective description through a 
density functional for $\rho$. This is a remarkable possibility because
it gets rid of the angular fields altogether which form the basis for
the definition of $\rho$ ! It may be understood by the fact that in
these configurations the angular fields are very rapidly fluctuating
functions and a spatial coarse graining procedure will eliminate them.
This is reminiscent of the fact that mean field or density functional
methods for nucleons in nuclear physics work very well without keeping
explicitly the dynamical pionic degrees of freedom. Similarly, in a
recent analysis of soliton formation in the Nambu-Jona Lasinio
model~\cite{ProvHans}
it was found that without enforcing the chiral circle condition 
stable minima are characterized by vanishing pion field.

Of course, it will be most interesting to extend the present
investigations to the $3D$-$O(4)$ model, where a lot of effort has 
gone into exploring $3D$-$SU(N_f)$ multi skyrmions~\cite{multisky} and
their possible relevance for the structure of nuclei. Their spatial
structure (which for large $B$ 
looks somehow like buckey balls) has little in common
with our naive picture of a nucleus. On the other hand it has been
known for a long time~\cite{sigma} that inclusion of a scalar $\sigma$ field
is important for the attractive part of the skyrmion-skyrmion force.
We should also rather expect a mechanism as described in sect.4 where the
nucleons keep their individuality inside the bag. For dimensional
reasons in the $3D$-$O(4)$ model there is no need for the Zeeman-type
coupling to monitor the scale of the structures; it emerges directly
from the competition between the second order (non linear $\sigma$) term
and the (fourth order) Skyrme term. Dominance of the Skyrme term
(which may arise in the interior of the bag due to the factor $\Phi^2$
multiplying the nonlinear $\sigma$-term,
or due to increasing temperature which lowers the coefficient of
the whole second order term) will again  
lead to deconfined baryon density inside the nucleus (or nuclear
matter in the limit $B\to \infty$). In any case, our present results
strongly suggest that for a description of nuclei in terms of 
multiply charged solitons in chiral meson fields the constraint
$\Phi \equiv 1$ is too restrictive.

Creation of spatial regions with disoriented chiral condensate (DCC) in
the course of a chiral symmetry-breaking transition has been studied in the
framework of the linear $O(4)$ model in trivial topology~\cite{DCC}. It
will be of 
interest to investigate how the existence of the various phase regions
explored here for the simple $2D$-$O(3)$ model may affect the
conclusions to be drawn for defect and DCC formation
in case of the chiral phase transition~\cite{Zapo}. 

Finally we should note that all results reported here have been
obtained through numerical relaxation of a lattice
functional through some Metropolis cooling algorithm. This naturally
poses the question about the continuum limit of these results.
Apparently, (as is directly obvious from fig.\ref{fig4}),
this is not a trivial matter because $B$ conservation relies
on the removal of one single point ($\Phi=0$) from the manifold on
which the model is defined. But this is the usual mathematical
difficulty encountered in the transition from a granular to
a continuous density distribution. And, again as usual, we have no
proof that we really have obtained the lowest energy minima in the
respective $B$-sectors.

\end{document}